\documentclass[conference]{IEEEtran}
\IEEEoverridecommandlockouts
\usepackage{cite}
\usepackage{amsmath,amssymb,amsfonts}
\usepackage{algorithmic}
\usepackage{graphicx}
\usepackage{textcomp}
\usepackage{xcolor}
\usepackage{svg}

\def\BibTeX{{\rm B\kern-.05em{\sc i\kern-.025em b}\kern-.08em
    T\kern-.1667em\lower.7ex\hbox{E}\kern-.125emX}}
\begin{document}

\title{HadAgent: Harness-Aware Decentralized \\
Agentic AI Serving with Proof-of-Inference \\
Blockchain Consensus}



\author{
\IEEEauthorblockN{Landy Jimenez\textsuperscript{1}, Mariah Weatherspoon\textsuperscript{1}, Bingyu Shen\textsuperscript{2}, Yi Sheng\textsuperscript{3}, Jianming Liu\textsuperscript{4},Boyang Li\textsuperscript{1}}
\IEEEauthorblockA{\textsuperscript{1}\textit{Department of Computer Science and Technology}, \\
\textit{Kean University},
Union, New Jersey, U.S.}
\IEEEauthorblockA{\textit{Department of Computer Science and Engineering},
\textit{University of Notre Dame}, \\
Notre Dame, Indiana, U.S. \\
}
\IEEEauthorblockA{\textsuperscript{3}\textit{Bellini College of Artificial Intelligence, Cybersecurity and Computing},
\textit{University of South Florida},\\
Tampa, Florida, U.S.}
\IEEEauthorblockA{\textsuperscript{4}\textit{DefinanceX, U.S.}}
{\tt\small \{jimenlan, weatherm, boli\}@kean.edu, bingyu.shen@hotmail.com, sheng1@usf.edu, liujm06@gmail.com}
}

\IEEEoverridecommandlockouts

\IEEEpubid{\makebox[\columnwidth]{978-8-3503-1019-1/23/\$31.00~\copyright2026 IEEE \hfill} \hspace{\columnsep}\makebox[\columnwidth]{ }}

\maketitle

\IEEEpubidadjcol

\begin{abstract}
Proof-of-Work (PoW) blockchain consensus consumes vast computational resources without producing useful output, while the rapid growth of large language model (LLM) agents has created unprecedented demand for GPU computation. We present HadAgent, a decentralized agentic AI serving system that replaces hash-based mining with Proof-of-Inference (PoI), a consensus mechanism in which nodes earn block-creation rights by executing deterministic LLM inference tasks. Because verification requires only re-executing a single forward pass under identical conditions, cross-node verification operates at consensus speed. HadAgent organizes validated records into a three-lane block body with dedicated DATA, MODEL, and PROOF channels, each protected by an independent Merkle root for fine-grained tamper detection. A two-tier node architecture classifies secondary nodes as trusted or non-trusted based on historical behavior: trusted nodes serve inference results in real time through optimistic execution, while non-trusted nodes must undergo full consensus verification. A harness layer monitors node behavior through heartbeat probes, anomaly detection via deterministic recomputation, and automated trust management, creating a self-correcting feedback loop that isolates malicious or unreliable participants. Experiments on a prototype implementation demonstrate 100\% detection rate and 0\% false positive rate for tampered records, sub-millisecond validation latency for record and hub operations, and effective harness convergence that excludes adversarial nodes within two rounds while promoting honest nodes to trusted status within five rounds.
\end{abstract}

\begin{IEEEkeywords}
Proof-of-Inference, Blockchain Consensus, Decentralized AI Inference, LLM Serving, Agent-to-Agent Interaction, Proof-of-Deep-Learning. 
\end{IEEEkeywords}

\section{Introduction} \label{sec:1_intro}

\begin{figure}[ht]
\centering
\includegraphics[width=.8\linewidth]{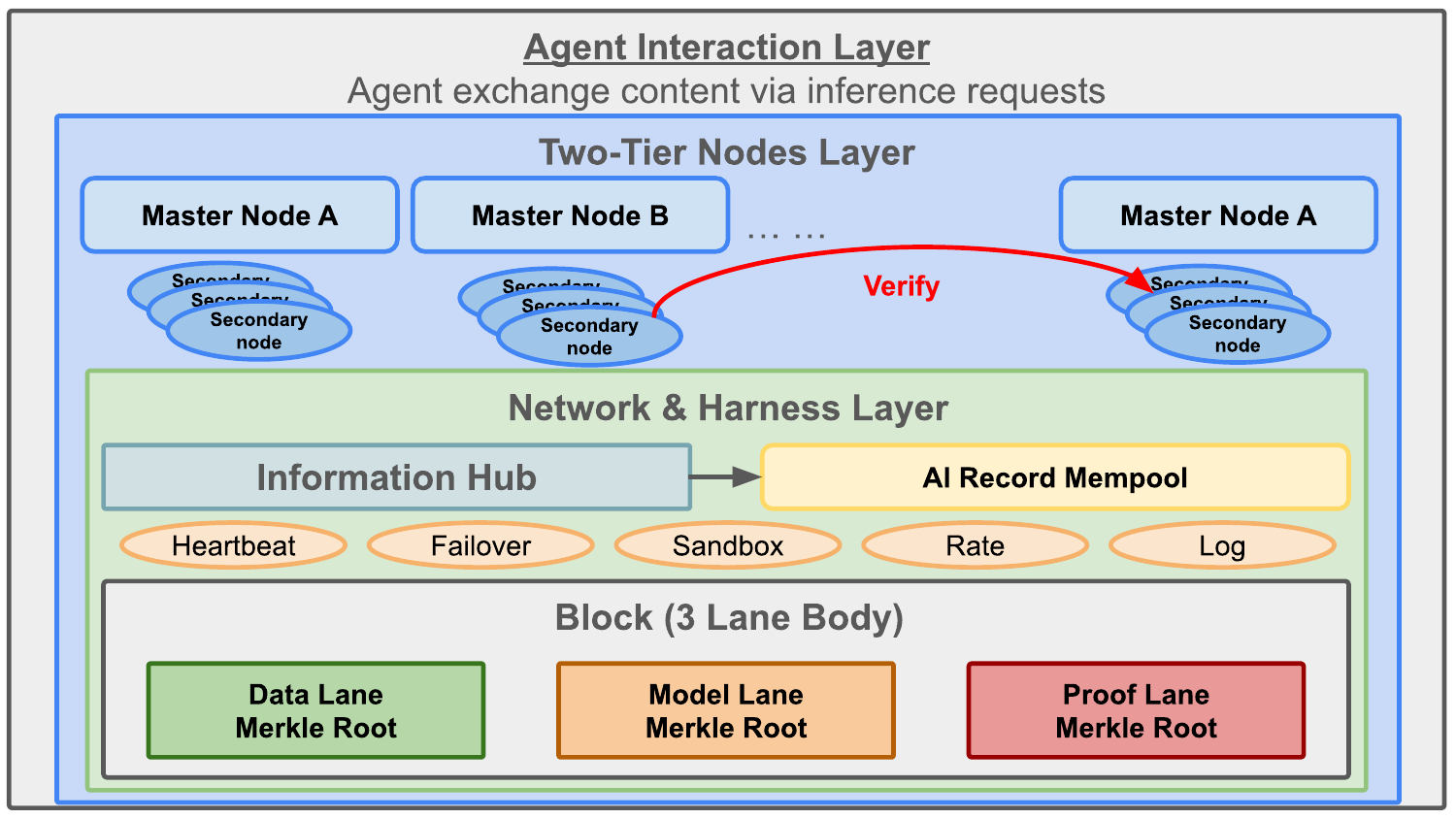}
\caption{Overview of the HadAgent system architecture. From top to bottom: the Agent Interaction Layer handles inference requests between agents; the Two-Tier Nodes Layer organizes master and secondary nodes with cross-node verification; the Network and Harness Layer provides communication, monitoring, and fault tolerance; and the Block layer stores validated records across three Merkle-rooted lanes (DATA, MODEL, PROOF). }\label{fig:overview}
\end{figure}

Blockchain systems have become a foundational technology for decentralized trust, yet the dominant consensus mechanism, Proof-of-Work
(PoW)~\cite{nakamoto2008bitcoin}, imposes a high computational cost with no productive output~\cite{8751419}. The Bitcoin network alone consumes over 100\, TWh of electricity annually~\cite{Bitcoin}, rivaling countries such as the Netherlands, while the computed hashes serve no purpose beyond block selection. At the same time, the rapid growth of large language models (LLMs) AI agents have created an unprecedented demand for GPU computation that consistently outpaces supply. This contradiction raises a natural question: can the computation required for blockchain consensus be redirected toward useful AI workloads?

Recent efforts have explored this direction by replacing hash-based mining with machine learning tasks. Systems such as Proof of Learning
(PoLe)~\cite{liu2021proof}, Coin.AI~\cite{baldominos2019coin}, and
DLBC~\cite{li2019dlbc} requires miners to train deep learning models on shared datasets, selecting the highest-accuracy node as the block producer. While these approaches demonstrate that meaningful computation can underpin consensus, they share three key limitations. First, model training may take hours or days, making it impractical for real-time services where inference responses are expected within seconds. Second, consensus is defined at the model level (who
trained the best model) rather than at the request level (whether a given output is correct for a given input), which does not match the granularity required for inference serving. Third, all nodes are treated uniformly regardless of their track record, introducing unnecessary verification latency and leaving no
mechanism to isolate unreliable participants.

We address these limitations in \textbf{HadAgent} (Harness-Aware Decentralized Agentic AI Serving) by making three design choices. First, we replace training-based consensus with \emph{Proof-of-Inference} (PoI): nodes execute deterministic inference tasks under controlled conditions (fixed weights, input, and decoding parameters) and submit the results as cryptographic proofs. Because verifying an inference output requires only a single forward pass rather than a full training run, cross-node verification is efficient. Second, we introduce a \emph{two-tier node architecture} that separates trusted nodes, which may serve requests in real time, from non-trusted nodes that must undergo full consensus verification. Third, a \emph{harness mechanism} monitors node behavior and intervenes upon anomalous patterns such as incorrect proofs or timeout violations, preventing cascading failures.

Figure~\ref{fig:overview} illustrates the layered architecture of HadAgent, comprising an Agent Interaction Layer, a Two-Tier Nodes Layer with master and secondary nodes, a Network and Harness Layer for communication and fault tolerance, and a three-lane Block body with independent Merkle roots for DATA, MODEL, and PROOF records.

In summary, this paper makes the following contributions:

\begin{itemize}

\item We propose Proof-of-Inference (PoI), a consensus mechanism in which the useful work is the AI service itself: nodes earn block-creation rights by executing deterministic LLM inference tasks whose results are independently verifiable through lightweight recomputation. Unlike prior proof-of-useful-work schemes that repurpose model training as mining, PoI directly couples consensus participation with real-time inference serving, ensuring that every unit of computation contributes to both network security and end-user value.

\item We design a three-lane block structure with dedicated DATA, MODEL, and PROOF channels, each protected by an independent Merkle root, supporting fine-grained tamper detection and auditable provenance.

\item We introduce a two-tier node architecture with a harness stability mechanism that enables real-time serving from trusted nodes while maintaining rigorous verification for non-trusted participants.

\end{itemize}






\section{Related Work} \label{sec:2_related} 
\subsection{Traditional Blockchain Consensus}
Blockchain consensus enables distributed nodes to agree on whether a new block should be added to the chain~\cite{nakamoto2008bitcoin}. Traditional blockchain systems mainly rely on Proof-of-Work (PoW) or Proof-of-Stake (PoS). PoW provides strong security by requiring nodes to solve computational puzzles, but it suffers from high energy consumption and computational waste~\cite{nakamoto2008bitcoin}. PoS improves efficiency by assigning consensus power according to financial stake rather than computation, although it may introduce challenges related to incentive design and centralization~\cite{king2012ppcoin}. These trade-offs have motivated continued research into more efficient and application-oriented consensus mechanisms~\cite{zheng2018blockchain}.

\subsection{Proof of Useful Work}
Recent work has explored replacing traditional Proof of Work (PoW) with useful computation, especially machine learning tasks, so that consensus can produce practical value rather than wasted hashes~\cite{11327514, 8751419, Li_2019_CVPR_Workshops, 10338941, 9461067, 10684436}. A representative direction is to use model training as the mining task. For example, Proof of Learning (PoLe), Coin.AI, and DLBC require nodes to train deep models and use model performance as the basis for consensus~\cite{liu2021proof, baldominos2019coin, li2019dlbc}. Other studies further extend this idea to distributed deep learning settings, where multiple nodes collaboratively contribute to useful computation while preserving fairness and verifiability~\cite{zhi2025blockchain}.

These works are grounded in the broader concept of Proof of Useful Work (PoUW), which argues that blockchain consensus should be tied to computations with real-world value, provided that the submitted work can be verified efficiently~\cite{haouari2022novel}. Existing blockchain literature also provides the foundation for secure block design, cryptographic integrity, and decentralized coordination~\cite{moosavi2024blockchain}. Our work follows this general direction, but shifts the focus from training-based useful work to deterministic AI inference, enabling lightweight verification and making the consensus process more suitable for real-time AI serving.

\subsection{Decentralized AI inference Network}
Several industry-scale projects have built decentralized networks for AI computation, each adopting a distinct trust model. Bittensor~\cite{rao2021bittensorpeertopeerintelligencemarket} constructs a peer-to-peer intelligence market where validators score miner outputs via the Yuma Consensus algorithm and distribute token emissions through stake-weighted aggregation, but its consensus is fundamentally subjective. Ritual~\cite{durvasula2025resonance, ritual2025chain} builds an AI-native execution layer that allows nodes to verify inference integrity through TEEs or zero-knowledge proofs, but TEE-based verification depends on hardware trust assumptions and ZKP-based verification for large models remains prohibitively expensive. TensorOpera (formerly FedML)~\cite{he2020fedml} demonstrates practical geo-distributed LLM serving across heterogeneous GPUs with master-worker routing, but operates as a centrally orchestrated platform with no on-chain verification. Gensyn~\cite{arun2025verdeverificationrefereeddelegation} introduces bitwise-deterministic ML primitives and refereed-delegation dispute resolution, but targets training workloads and treats verification as a dispute layer rather than a consensus mechanism.

\subsection{Harness Engineering}
Recent work on harness engineering argues that the reliability of agentic AI systems depends not only on model capability, but also on the surrounding orchestration, observability, and feedback mechanisms \cite{wu2026basicsletconversationalagents}. Our system follows this view by introducing a harness layer for monitoring, anomaly detection, and trust management in decentralized inference networks.

\textbf{Our work} occupies a distinct position along this spectrum. Unlike Bittensor's subjective scoring, we require every inference result to be independently reproducible through deterministic recomputation. Unlike Ritual, we avoid both hardware trust assumptions and expensive cryptographic proofs by exploiting the execution--verification cost asymmetry inherent in deterministic inference. We inherit TensorOpera's insight that a master node should route requests to distributed workers, but embed this topology within a blockchain consensus framework with cross-node verification. And while we share Gensyn's commitment to deterministic execution, we focus on inference serving and use deterministic evaluation as the consensus mechanism itself, directly coupling block creation rights with the production of verifiable inference results.


\section{Design} \label{sec:4_design}

\subsection{System Overview}

\subsubsection{Design Goals}


The system is designed to replace traditional Proof-of-Work (PoW) with meaningful AI computation, enabling nodes to contribute useful work, specifically large language model inference, while participating in consensus. The primary objective is deterministic evaluation: given identical model weights, input data, and decoding parameters, every compliant node must produce bit-identical output, allowing any submitted proof to be independently verified through lightweight recomputation of a single forward pass rather than a full training run.
 
Beyond correctness, the design targets three operational goals. First, real-time serving: trusted nodes should deliver inference results immediately through optimistic execution, without waiting for consensus. Second, security and tamper resistance: digital signatures and three-lane Merkle roots must ensure that all records are authenticated and that any modification is detectable. Third, fault tolerance: the harness mechanism must detect and contain anomalous node behavior such as missed deadlines, incorrect proofs, or node failures before it affects consensus progress or service quality.

\subsubsection{Architecture Overview} 
As illustrated in Figure~\ref{fig:overview}, HadAgent is organized into four layers. The \textbf{Agent Interaction Layer} serves as the entry point where external agents submit inference requests specifying an input prompt, model reference, and decoding parameters. The \textbf{Two-Tier Nodes Layer} organizes participants into \textbf{master nodes}, which route tasks, evaluate results, and propose blocks, and \textbf{secondary nodes}, which execute inference computation. Secondary nodes are further classified as \textbf{trusted} or \textbf{non-trusted (new)} based on historical behavior: trusted nodes may return results immediately through optimistic execution, while non-trusted nodes must wait for full consensus verification (Figure~\ref{fig:node}). The \textbf{Network and Harness Layer} contains the \textbf{Information Hub} for message routing, the \textbf{AI Record Mempool} for staging validated records, and harness services including heartbeat monitoring, failover, sandbox isolation, rate limiting, and logging. The \textbf{Block Layer} commits validated records to a three-lane block body (DATA, MODEL, PROOF), each protected by an independent Merkle root (Figure~\ref{fig:block}); only hashes, metadata, scores, and signatures are stored on-chain.
 
A complete inference cycle proceeds as follows (Figure~\ref{fig:flow}): (1) an agent submits a request; (2) a master node routes the task to secondary nodes; (3) secondary nodes execute deterministic inference and return results; (4) the master node evaluates the results and generates DATA, MODEL, and PROOF records; (5) records enter the mempool after schema and signature validation; (6) multiple master nodes participate in cross-master voting; (7) agreed-upon records are committed to a new block; and (8) the validated response is delivered to the requesting agent.

\subsubsection{Threat Model} 

We adopt a Byzantine fault model in which up to $f$ out of $N$ nodes may deviate arbitrarily from the protocol.
 
\textbf{Fabricated inference results.} A malicious secondary node may submit a forged PROOF record with an incorrect score or substituted hashes. Because inference is deterministic, any honest node can detect such deviations by re-executing the same task and comparing the output. This defense is realized by the Proof-of-Inference consensus mechanism described in Section~\ref{sec:poi}.
 
\textbf{Dishonest master nodes.} A compromised master node may discard valid proofs or inflate scores for a colluding secondary node. HadAgent mitigates this through cross-master verification: block proposals require agreement from multiple master nodes via multi-master voting, so no single master can unilaterally determine block contents. The voting procedure is detailed in Section~\ref{sec:poi}.
 
\textbf{Sybil attacks.} An adversary may create a large number of secondary nodes to dominate consensus. The two-tier architecture described in Section~\ref{sec:twotier} limits this threat: newly joined nodes are classified as non-trusted and must undergo full verified execution for every task, so Sybil nodes cannot bypass verification and must invest real computation for each proof.
 
We note that network-level adversaries capable of message delay, reordering, or selective dropping may disrupt consensus progress. The harness layer (Section~\ref{sec:harness}) provides basic mitigation through heartbeat monitoring and failover, but our current prototype does not fully address threats such as eclipse attacks. Hardening the communication layer with a production-grade P2P protocol is left as future work (Section~\ref{sec:6_disc}).

\subsubsection{Assumptions} 
The correctness and security guarantees of HadAgent rely on three core assumptions.
 
Assumptions 1 (Deterministic inference): Given identical model weights, input data, and decoding parameters, every compliant node produces identical output. This allows any node to verify a submitted proof by re-executing the same task and comparing results. We enforce this by requiring all nodes to use a standardized execution environment with a fixed framework version and numerical precision.
 
Assumptions 2 (Honest majority among master nodes): Strictly more than half of the master nodes follow the protocol faithfully. Under this assumption, cross-master voting produces correct consensus decisions, and no coalition of dishonest masters can approve fabricated proofs or censor valid ones.
 
Assumptions 3 (Model and data availability): All nodes participating in a given inference task can retrieve the same model weights and evaluation dataset through an off-chain distribution channel. On-chain content hashes allow every node to verify that the artifacts it has obtained match those used by other participants.
 
Additional practical considerations, including bounded clock drift, cryptographic hardness, and dynamic node membership, are discussed as limitations in Section~\ref{sec:6_disc}.

\begin{figure}[!t]
\centering
\includegraphics[width=.85\linewidth]{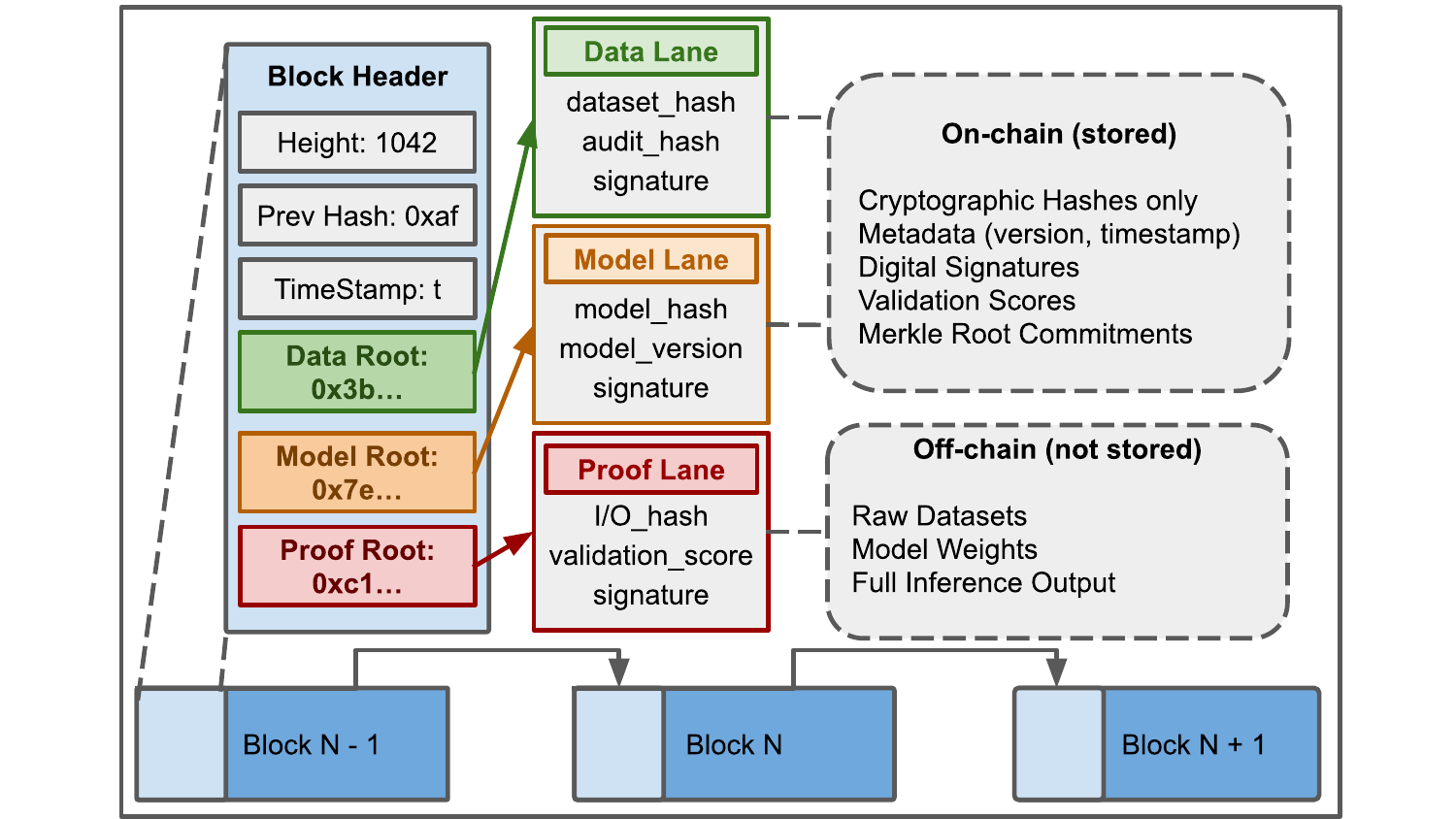}
\caption{Overview of Block Design. Each block contains a header and a three-lane body. The header stores the block height, previous block hash, timestamp, and three independent Merkle roots corresponding to the DATA, MODEL, and PROOF lanes. The Data Lane records dataset and audit hashes, the Model Lane records model hashes and version identifiers, and the Proof Lane records input/output hashes and validation scores; all lanes include digital signatures for authentication. Only cryptographic hashes, metadata, validation scores, and signatures are stored on-chain; raw datasets, model weights, and full inference outputs remain off-chain and are referenced by their content hashes.}
\label{fig:block}
\end{figure}

\subsection{Block Structure and Data Model} 
Figure~\ref{fig:block} shows the overview of the block design. Each block consists of a header and a body. The header stores metadata required for validation, including the block height, previous block hash, and root hashes that summarize the contents of the block. These root hashes allow nodes to verify integrity without processing all records.

The block body contains validated records collected from the mempool. Records are organized by type and contribute to the computation of their corresponding root hashes. This structure enables efficient verification while maintaining a clear separation between different categories of data. 

\subsubsection{Three Lane Block Body} 
Figure~\ref{fig:overview} illustrates the layered architecture of HadAgent and the three-lane block body, with Figure~\ref{fig:block} providing the detailed schema of each lane. The block body is organized into three logical lanes: DATA, MODEL, and PROOF. Each lane contains records of a specific type, allowing independent processing and validation. The DATA lane stores data commitments, the MODEL lane contains model references, and the PROOF lane records evaluation results.

This separation simplifies validation by separating different responsibilities within the block. It also supports root computation, as each lane contributes to a unique hash in the block header. 

\subsubsection{Record Types} 
The system defines three record types: DATA, MODEL, and PROOF. Each type captures a different part of the AI-based consensus workflow and follows a structured schema to support validation and secure storage.

A DATA record is used to represent a commitment to an input dataset or data artifact. Its fields include a content hash, a timestamp, an owner or sender identifier, optional metadata, and a digital signature. The content hash uniquely identifies the off-chain data and allows other nodes to verify its integrity without storing the raw data on-chain.

A MODEL record is used to describe a model configuration or model artifact associated with an inference task. Its fields include \texttt{model\_hash}, \texttt{model\_version}, a model identifier, configuration metadata, a timestamp, and a digital signature. The \texttt{model\_hash} binds the record to a specific model configuration or serialized model artifact, while \texttt{model\_version} distinguishes between different revisions of the same model.

A PROOF record stores the result of an evaluation run and serves as the main evidence used in consensus. Its fields include \texttt{dataset\_hash}, \texttt{model\_hash}, \texttt{validation\_score}, task or proof identifiers, a timestamp, and a digital signature. The \texttt{dataset\_hash} links the proof to the exact evaluation dataset, while the \texttt{validation\_score} records the outcome of deterministic inference under the agreed configuration.

Before any record is accepted, it undergoes schema validation. This process checks that all required fields are present, that each field matches the expected type and format, and that fixed-length values such as hashes satisfy size limits. For example, cryptographic hashes must have the correct byte length, version fields must be valid identifiers, and scores must be encoded in the expected numeric format. Records that fail these structural checks are rejected immediately.

After schema validation, the system performs signature verification. The record payload is processed into a consistent format, and the included digital signature is verified against the sender's public key. This step confirms both authenticity and integrity: it proves which node created the record and ensures that the contents were not modified after signing. Only records that pass both schema validation and signature verification are admitted into the mempool and considered for inclusion in a block.

\subsubsection{On-chain vs Off-chain Storage}
\label{sec_on_off_chain}


Blockchain has different storage methods within itself to determine how it wants to store data. Each offers different benefits in terms of what data set it corresponds with. This concept of blockchain is exceptionally important because it helps support the whole base of the blockchain itself. It ensures all data is being kept secure and properly transmitted~\cite{Li_2019_CVPR_Workshops}.

To start, on-chain methods ensure immutability and reliability by storing data directly on the blockchain, although this carries its limitations. The main concern is its transactional cost and lack of scalability~\cite{Li_2019_CVPR_Workshops}. Off-chain methods, on the other hand, offer better cost and performance advantages but run the risk related to data integrity. Off-chain stores the data outside the blockchain system while keeping the data references or the cryptographic hashes on the chain.'

With Proof-of-Inference (PoI), it takes out the need for raw data logistics. We can look at an example of a supply chain, where data such as location throughout the shipment is collected, but this type of data is stored off of the chain. Machine learning analyzes the data, and instead of recording all the logistical data, PoI only stores the verification result. We can compare this similarly to off-chain, although PoI does it more efficiently.

\subsection{Network and Communication} 
The system includes a lightweight communication layer that enables nodes to exchange records and validation messages. The design focuses on simplicity and supports controlled experimentation while maintaining compatibility with decentralized architectures. 

\subsubsection{Information Hub Layer}
The information hub acts as the entry point for incoming messages. It processes structured packets and routes them to appropriate components such as validation or the mempool. Communication is implemented using socket-based messaging.

The hub operates asynchronously, allowing nodes to handle multiple requests concurrently. Incoming messages are validated before being processed to ensure correctness.

\subsubsection{Record Mempool}
The mempool temporarily stores validated records before they are included in a block. It maintains separate collections for different record types and enforces validation before insertion.

The mempool also prevents duplicate entries and supports efficient selection of records during block creation. This ensures that only valid records are propagated through the system.

\subsubsection{P2P Gossip and Peer Discovery}

In a fully decentralized deployment, nodes would rely on a P2P gossip protocol with automatic peer discovery to propagate data, model, and proof records across the network.
Getting this right is a research problem in its own right, as NAT traversal, Sybil-resilient peer selection, and efficient gossip under node churn are each non-trivial and largely separate from the consensus mechanism we focus on here.
Moreover, at the scale of our testbed (X nodes), the benefits of a production-grade gossip layer are minimal: every node can maintain direct connections to every other node without bandwidth or connectivity concerns.

We therefore adopt a simplified communication design in which nodes exchange typed packets \texttt{NEW\_DATA\_RECORD}, \texttt{NEW\_MODEL\_RECORD}, \texttt{NEW\_PROOF\_RECORD} over point-to-point TCP connections using Python's asyncio-based networking. Each node connects to a known set of peers, which is sufficient for evaluating the correctness and performance of the consensus mechanism itself. We acknowledge that this direct-socket architecture does not capture the latency variability and message loss patterns of a real-world P2P overlay; a full integration with an established gossip protocol (e.g., libp2p) remains future work.

\begin{figure}
\centering
\includegraphics[width=.85\linewidth]{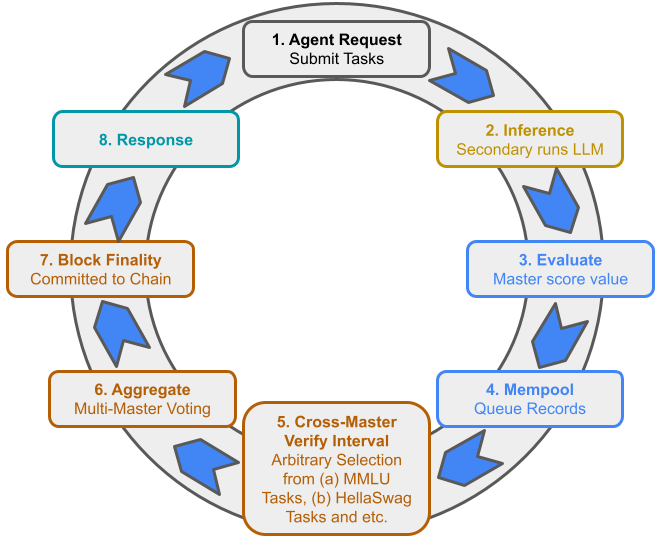}
\caption{Consensus Flow with Interval Verification. The cycle proceeds in eight steps: (1)~an agent submits an inference request; (2)~a secondary node executes the LLM inference task; (3)~the master node evaluates the returned result and computes a validation score; (4)~the validated record is queued in the mempool; (5)~a cross-master verification interval is triggered, in which master nodes independently re-evaluate a randomly selected subset of tasks drawn from standardized benchmarks (e.g., MMLU, HellaSwag); (6)~multiple master nodes aggregate their verification outcomes through multi-master voting; (7)~the agreed-upon records are committed to the chain as a finalized block; and (8)~the validated response is returned to the requesting agent.}
\label{fig:flow}
\end{figure}

\subsection{Proof-of-Inference Consensus} 
\label{sec:poi}

In Proof-of-Inference (PoI), nodes earn block-creation rights by executing deterministic LLM inference tasks whose correctness can be verified through a single forward pass recomputation. The consensus flow proceeds in eight steps as shown in Figure~\ref{fig:flow}. An agent submits an inference request (step 1), and a master node assigns the task to one or more secondary nodes. Each secondary node executes the inference under fixed model weights, input data, and decoding parameters, producing a deterministic output (step 2). The master node evaluates the result by recomputing the same forward pass and generates a PROOF record containing the dataset hash, model hash, validation score, and digital signature (step 3), which is then queued in the mempool (step 4).
 
Periodically, a cross-master verification interval is triggered (step 5), during which master nodes independently re-evaluate a randomly selected subset of recent tasks drawn from standardized benchmarks such as MMLU and HellaSwag. The random selection prevents secondary nodes from predicting which tasks will be audited. Master nodes then aggregate their verification outcomes through master voting (step 6). If voting confirms consistency across independent recomputations, the records are committed to the chain as a finalized block (step 7) and the response is delivered to the agent (step 8).

\subsection{Inference Task Lifecycle} 
The inference task lifecycle describes how evaluation tasks are executed, validated, and recorded within the system. This process replaces traditional mining by requiring nodes to perform deterministic model evaluation and submit the results as proofs.

The lifecycle begins when a node prepares an evaluation task using a shared dataset and a predefined model configuration. The dataset is referenced by its cryptographic hash to ensure consistency across all nodes. The node then executes the model under fixed conditions, producing outputs that are deterministic and reproducible. 

After execution, the node computes a validation score by comparing the model outputs to expected results. This score represents the performance of the model for the given task. The node then generates a PROOF record containing the dataset hash, model configuration reference, validation score, and a digital signature.

The proof record is submitted to the system through the information hub, where it undergoes validation. Other nodes verify the proof by recomputing the evaluation and checking that the results match the submitted score. Signature verification is also performed to ensure authenticity.

Once validated, the record is stored in the mempool and later included in a block during block creation. This lifecycle ensures that all inference tasks are reproducible, verifiable, and consistently integrated into the blockchain.

\begin{figure}[ht]
\centering
\includegraphics[width=\linewidth]{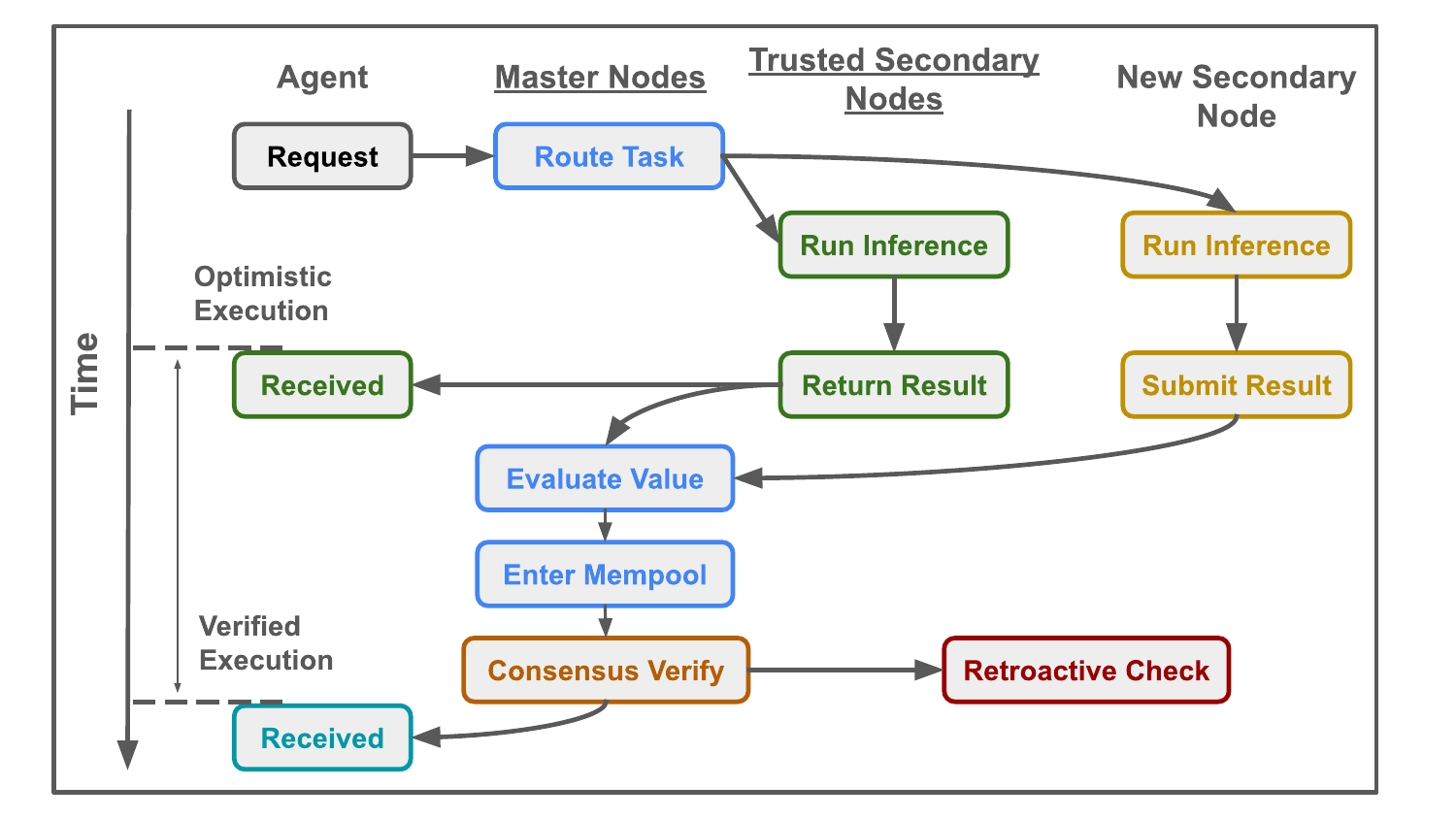}
\caption{Demonstration of the Two-Tier Node Architecture and Inference Paths. \textbf{Left (Optimistic Execution):} When a trusted secondary node handles a request, the master node routes the task, the trusted node runs inference and returns the result directly to the agent without waiting for consensus, enabling real-time serving. The master node evaluates the result and enters it into the mempool in the background. \textbf{Right (Verified Execution):} When a new (non-trusted) secondary node handles a request, its result is not delivered immediately. Instead, the submitted result must pass master evaluation, enter the mempool, and undergo full consensus verification with a retroactive check before the response is delivered to the agent.}
\label{fig:node}
\end{figure}

\subsection{Two-Tier Node Architecture} \label{sec:twotier}
 
HadAgent classifies secondary nodes into two tiers based on their historical behavior, as illustrated in Figure~\ref{fig:node}. A \textbf{trusted secondary node} is one that has accumulated a sufficient record of correct proofs without anomalies. When a trusted node receives a task from the master, it executes inference and returns the result directly to the requesting agent through \textbf{optimistic execution}, enabling real-time serving. The master node still evaluates the result and enters the corresponding records into the mempool in the background, but the agent does not wait for consensus.
 
A \textbf{non-trusted secondary node} is either newly joined or has been demoted due to past failures. Its inference result is not delivered to the agent immediately. Instead, the result must pass master evaluation, enter the mempool, and undergo full consensus verification before it is accepted. This \textbf{verified execution} path ensures that unproven nodes cannot inject incorrect results into the system.
 
Promotion from non-trusted to trusted status requires a node to complete a consecutive sequence of verified tasks without any rejection. Demotion occurs when the harness layer (Section~\ref{sec:harness}) detects anomalous behavior such as incorrect proofs or repeated timeouts. This creates a natural cost asymmetry: earning trust requires sustained correct behavior, while losing it requires only a single verified failure.

\subsection{Harness Engineering for System Stability} \label{sec:harness}
 
The harness layer continuously monitors node behavior and enforces the trust transitions described in Section~\ref{sec:twotier}. It operates in three sequential phases at the end of each consensus round.
 
In the first phase, a heartbeat monitor sends liveness probes to all secondary nodes. Any node that fails to respond within a configurable timeout is flagged as unresponsive and its pending tasks are reassigned to available nodes.
 
In the second phase, an anomaly detector verifies every proof record submitted during the round. For each record, the detector recomputes the expected validation score by re-executing the same deterministic inference task and compares it against the submitted score. A deviation exceeding a predefined threshold indicates a fabricated or degraded result and is flagged as an anomaly.
 
In the third phase, a trust manager updates each node's tier based on the accumulated evidence. A node that triggers consecutive anomalies or timeouts beyond a demotion threshold (e.g., 2 consecutive failures) is demoted from trusted to non-trusted, or from non-trusted to excluded. Conversely, a non-trusted node that completes a sufficient number of consecutive correct proofs (e.g., 5 consecutive rounds without rejection) is promoted to trusted. Excluded nodes are removed from task assignment entirely.
 
This three-phase pipeline creates a self-correcting feedback loop: malicious nodes that submit inflated scores are detected through deterministic recomputation and progressively isolated, while honest nodes that maintain consistent behavior are gradually promoted to enable real-time serving. We implement a prototype of this pipeline using CrewAI, where each phase is modeled as an autonomous agent (HeartbeatMonitor, AnomalyDetector, TrustManager) that executes sequentially within each round.

\section{Experiment} \label{sec:5_exper} 

\subsection{Experiment Setup} 
The experiments were conducted using a Python-based prototype implementation of the proposed blockchain system. The system includes components for block validation, record validation, an information hub for message handling, and a mempool for temporary storage of validated records. 

All experiments were executed on a local machine running macOS, using Python 3.14 in a virtual environment. The test suite was implemented using \texttt{pytest}, enabling automated execution of test cases and validation of expected outcomes. Asynchronous components of the system were tested using \texttt{asyncio}-based execution.

The evaluation was divided into three phases: (1) a baseline correctness evaluation, (2) a scale evaluation, and (3) a combined evaluation. The baseline phase verifies correctness using targeted valid and invalid inputs, including tampered records and blocks. The scale phase evaluates system performance by processing a large number of valid records to simulate realistic workload conditions. The combined phase integrates both baseline and scale tests into a single run, allowing simultaneous evaluation of correctness, rejection accuracy, and performance under load.

During each test, timestamps were recorded at the start and end of validation using high-resolution timers, allowing measurement of latency for individual operations. Metrics collected include the number of valid and invalid cases, detection rate, false positive rate, and validation latency. All results were logged and exported as structured JSON files for analysis.

All experiments were conducted in a single-node testing environment without distributed networking, focusing on validating system logic, correctness, and performance under controlled conditions.

\subsection{Block Integrity Tests} 



To evaluate the accuracy of the block structure, a series of integrity tests were conducted focusing on structural and cryptographic properties of the blockchain. 

Block height validation was tested by constructing blocks with incorrect height values and verifying that the system correctly rejects them. Previous hash verification was evaluated by creating blocks with invalid references to prior blocks, ensuring that incorrect chain linkage is detected.

Additionally, root hash validation was performed by modifying the data root, model root, and proof root values in the block header. These tests confirm that any tampering with block contents is reflected in the corresponding root hash and detected during validation.

Signature validation was also tested by replacing valid block signatures with invalid values. The system correctly rejected blocks with forged or malformed signatures, confirming proper authentication enforcement.

Finally, chain linkage correctness was verified by constructing sequential blocks with valid parameters and confirming that properly linked blocks are accepted. 

The results demonstrate that the system enforces strict validation across block structure, linkage, integrity, and authentication, preventing tampered or invalid blocks from being accepted.

\subsection{Proof Verification Accuracy} 


To evaluate proof verification accuracy, tests were conducted using both valid and intentionally malformed records. These tests simulate scenarios where records contain missing fields, incorrect hash lengths, or invalid digital signatures.

The system performs schema validation to ensure all required fields are present and correctly formatted. It then verifies cryptographic signatures to confirm authenticity and integrity. Records that fail either step are rejected before entering the mempool.

\begin{table}[h]
\centering
\caption{Proof Verification Accuracy Results}
\label{tab:accuracy}
\begin{tabular}{lcc}
\hline
\textbf{Metric} & \textbf{Baseline} & \textbf{Combined} \\
\hline
\textit{Records} \\
\quad Valid Records Tested & 3 & 1003 \\
\quad Invalid Records Tested & 8 & 8 \\

\textit{Blocks} \\
\quad Valid Blocks Tested & 2 & 2 \\
\quad Invalid Blocks Tested & 5 & 5 \\

\textit{Hub} \\
\quad Valid Hub Tests & 5 & 1005 \\
\quad Invalid Hub Tests & 2 & 2 \\

\textit{Mempool} \\
\quad Valid Pool Tests & 3 & 3 \\
\quad Invalid Pool Tests & 1 & 1 \\

\hline
Total Valid Cases & 13 & 2013 \\
Total Invalid Cases & 16 & 16 \\
\hline
\textbf{Detection Rate} & \textbf{100\%} & \textbf{100\%} \\
\textbf{False Positive Rate} & \textbf{0\%} & \textbf{0\%} \\
\hline
\end{tabular}
\end{table}

In the baseline evaluation, a total of 16 invalid cases were introduced, including schema violations and signature tampering. All invalid cases were correctly rejected, resulting in a detection rate of 100\%. Additionally, no valid records were incorrectly rejected, yielding a false positive rate of 0\%.

In the combined evaluation, the same validation logic was applied alongside large-scale record processing. The system maintained a 100\% detection rate and 0\% false positive rate while handling over 2000 validation operations.

These results demonstrate that the system can accurately distinguish valid records from tampered or malformed inputs, ensuring reliable proof verification within the consensus process.

\subsection{Consensus Latency} 



\begin{figure}
\centering
\includegraphics[width=\linewidth]{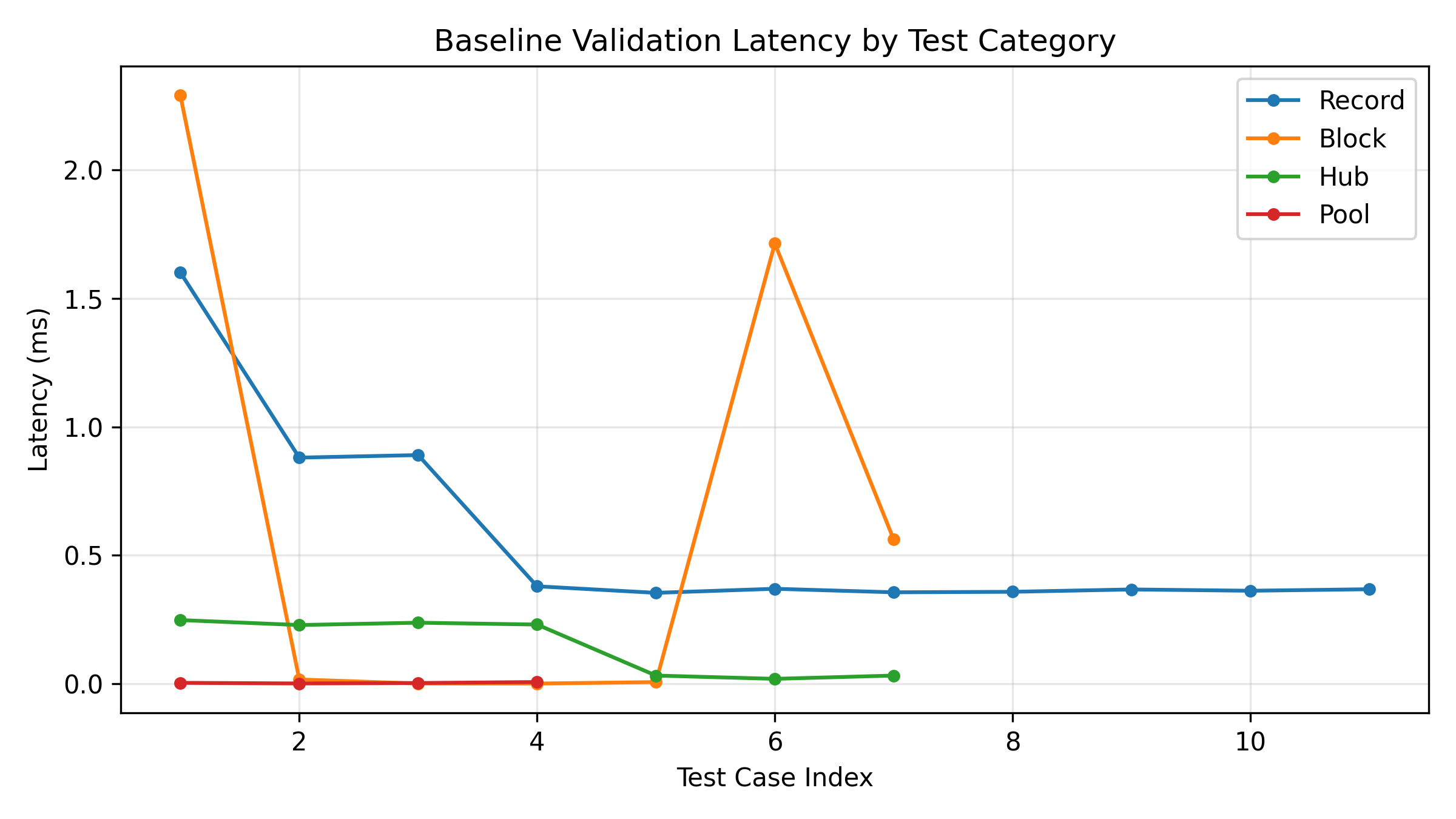}
\caption{Consensus Latency Performance}
\label{fig:baseline_latency}
\end{figure}

To evaluate system performance, latency was measured as the time between record submission and validation result within the consensus pipeline. This corresponds to the validation phase of consensus, where records and blocks are verified before being accepted.

Timestamps were recorded at the start and end of each validation operation using high-resolution timers. Latency was computed as the difference between these timestamps, allowing precise measurement of processing time for individual operations.

Figure~\ref{fig:baseline_latency} shows per-test validation latency across system components in the baseline evaluation. Record and hub operations exhibit consistently low latency with minimal variation, generally remaining below 1 ms. Block validation incurs higher latency and occasional spikes due to additional structural and cryptographic checks. Pool operations exhibit negligible latency and are included for completeness.

In the scale evaluation, the system processed over 2000 validation operations, including 1000 record validations and 1000 hub submissions. Latency remained consistent with baseline measurements, indicating stable performance under increased workload. Latency measurements exhibit low variance for record and hub operations, while block validation shows expected variability.

These results demonstrate that the system maintains low-latency validation while scaling efficiently. It is important to note that these measurements capture the validation component of the consensus process rather than full end-to-end latency from inference request submission to block confirmation.


\subsection{Harness Engineering for System Stability} \label{exp_secs:harness}
 
The harness layer wraps around the inference pipeline to detect, contain, and permanently correct anomalous node behavior. Following the harness engineering paradigm~\cite{hashimoto2026harness}, every detected failure is not merely rejected but used to tighten constraints on the offending node, ensuring the same failure cannot recur in subsequent rounds.
 
The harness operates in three sequential phases at the end of each consensus round. In the heartbeat phase, the harness sends liveness probes to all secondary nodes. A node that fails to respond within a configurable timeout is flagged as unresponsive, and its pending tasks are immediately reassigned to an available secondary node through the failover mechanism. This prevents a single unresponsive node from stalling the consensus pipeline.
 
In the anomaly detection phase, the harness verifies every proof record submitted during the round. For each PROOF record, the harness re-executes the same inference task under identical conditions (model weights, input data, decoding parameters) and compares the recomputed validation score against the submitted value. If the absolute deviation exceeds a predefined tolerance threshold, the record is flagged as anomalous. Because inference is deterministic (Assumption A1), any non-zero deviation beyond numerical tolerance indicates either fabrication or corruption.
 
In the trust update phase, the harness applies tier transition rules to each node based on the accumulated evidence from the preceding two phases. A node that accumulates consecutive failures (anomalies or timeouts) exceeding a demotion threshold $\tau_d$ is downgraded: from trusted to non-trusted, or from non-trusted to excluded. An excluded node is removed from the task assignment pool entirely. Conversely, a non-trusted node that completes consecutive correct proofs exceeding a promotion threshold $\tau_p$ is upgraded to trusted, enabling it to serve future requests through optimistic execution (Section~\ref{sec:twotier}). In our prototype, we set $\tau_d = 2$ and $\tau_p = 5$, reflecting the design principle that losing trust should be fast while earning trust should be gradual.
 
This three-phase pipeline creates a self-correcting feedback loop between the harness layer and the two-tier node architecture. Malicious nodes that submit inflated scores are detected through deterministic recomputation and progressively isolated within two rounds. Unreliable nodes that repeatedly time out are excluded through the same demotion path. Meanwhile, honest nodes that maintain consistent behavior are gradually promoted, increasing the proportion of trusted nodes capable of real-time serving.

\section{Limitation and Future Work} \label{sec:6_disc}

\textbf{Production-grade P2P networking:} Our current prototype relies on direct TCP connections among a pre-configured set of peers. A natural next step is to replace this layer with a structured gossip protocol such as \texttt{libp2p}, enabling automatic peer discovery, NAT traversal, and resilient message propagation across larger and more dynamic networks. This would also allow us to study how real-world network conditions, including packet loss, variable latency, and node churn, affect consensus convergence and proof propagation time.

\textbf{Additional assumptions:} The current design assumes bounded clock drift for timeout-based harness mechanisms, standard cryptographic hardness for SHA-256 and ECDSA, and a fixed node set without dynamic membership changes. Additionally, our experiments use a homogeneous execution environment; achieving bit-level deterministic inference across heterogeneous GPU hardware may require techniques such as software-emulated floating-point arithmetic. We plan to address these limitations in future work.

Overall, these limitations do not change the central contribution of this work: demonstrating that deterministic AI inference can serve as a useful and verifiable consensus mechanism. We hope this prototype provides a foundation for future research on scalable, trustworthy, and real-time decentralized AI serving.














\clearpage
\bibliographystyle{IEEEtran} 
\bibliography{bib}

@article{nakamoto2008bitcoin,
  title={Bitcoin: A peer-to-peer electronic cash system},
  author={Nakamoto, Satoshi and others},
  year={2008},
  publisher={Working Paper}
}

@article{king2012ppcoin,
  title={Ppcoin: Peer-to-peer crypto-currency with proof-of-stake},
  author={King, Sunny and Nadal, Scott},
  journal={self-published paper, August},
  volume={19},
  number={1},
  year={2012}
}

@INPROCEEDINGS{11327514,
  author={Wen, Jinghao and Ma, Dongning and Zhang, Sizhe and Sudler, Hasshi and Jiao, Xun},
  booktitle={2025 IEEE Biomedical Circuits and Systems Conference (BioCAS)}, 
  title={Proof-of-Useful-Work Blockchain for Trustworthy Biomedical Hyperdimensional Computing}, 
  year={2025},
  volume={},
  number={},
  pages={56-60},
  doi={10.1109/BioCAS67066.2025.00023}}

@INPROCEEDINGS{10338941,
  author={Li, Boyang and Shen, Bingyu and Lu, Qing and Jung, Taeho and Shi, Yiyu},
  booktitle={2023 Fifth International Conference on Blockchain Computing and Applications (BCCA)}, 
  title={Proof-of-Federated-Learning-Subchain: Free Partner Selection Subchain Based on Federated Learning}, 
  year={2023},
  volume={},
  number={},
  pages={600-605},
  doi={10.1109/BCCA58897.2023.10338941}}

@INPROCEEDINGS{9461067,
  author={Li, Boyang and Lu, Qing and Jiang, Weiwen and Jung, Taeho and Shi, Yiyu},
  booktitle={2021 IEEE International Conference on Blockchain and Cryptocurrency (ICBC)}, 
  title={A mining pool solution for novel proof-of-neural-architecture consensus}, 
  year={2021},
  volume={},
  number={},
  pages={1-3},
  doi={10.1109/ICBC51069.2021.9461067}}

@ARTICLE{10684436,
  author={Qu, Xidi and Wang, Shengling and Li, Kun and Huang, Jianhui and Cheng, Xiuzhen},
  journal={IEEE Transactions on Mobile Computing}, 
  title={TidyBlock: A Novel Consensus Mechanism for DAG-based Blockchain in IoT}, 
  year={2025},
  volume={24},
  number={2},
  pages={722-735},
  doi={10.1109/TMC.2024.3464494}}

@InProceedings{Li_2019_CVPR_Workshops,
author = {Li, Boyang and Chenli, Changhao and Xu, Xiaowei and Jung, Taeho and Shi, Yiyu},
title = {Exploiting Computation Power of Blockchain for Biomedical Image Segmentation},
booktitle = {Proceedings of the IEEE/CVF Conference on Computer Vision and Pattern Recognition (CVPR) Workshops},
month = {June},
year = {2019}
}

@misc{hashimoto2026harness,
  author = {Mitchell Hashimoto},
  title = {My {AI} Adoption Journey},
  year = {2026},
  url = {https://mitchellh.com/writing/my-ai-adoption-journey},
}

@misc{wu2026basicsletconversationalagents,
      title={Back to Basics: Let Conversational Agents Remember with Just Retrieval and Generation}, 
      author={Yuqian Wu and Wei Chen and Zhengjun Huang and Junle Chen and Qingxiang Liu and Kai Wang and Xiaofang Zhou and Yuxuan Liang},
      year={2026},
      eprint={2604.11628},
      archivePrefix={arXiv},
      primaryClass={cs.CL},
      url={https://arxiv.org/abs/2604.11628}, 
}

@misc{Bitcoin,
    author = {Digiconomist},
    title = {Bitcoin Energy Consumption Index {@ONLINE}},
    month = {March},
    year = {2019},
    howpublished={\url{https://digiconomist.net/bitcoin-energy-consumption}},
    note = "(accessed: 03.06.2019)",
}

@article{he2020fedml,
  title={Fedml: A research library and benchmark for federated machine learning},
  author={He, Chaoyang and Li, Songze and So, Jinhyun and Zeng, Xiao and Zhang, Mi and Wang, Hongyi and Wang, Xiaoyang and Vepakomma, Praneeth and Singh, Abhishek and Qiu, Hang and others},
  journal={arXiv preprint arXiv:2007.13518},
  year={2020}
}

@misc{rao2021bittensorpeertopeerintelligencemarket,
      title={BitTensor: A Peer-to-Peer Intelligence Market}, 
      author={Yuma Rao and Jacob Steeves and Ala Shaabana and Daniel Attevelt and Matthew McAteer},
      year={2021},
      eprint={2003.03917},
      archivePrefix={arXiv},
      primaryClass={cs.AI},
      url={https://arxiv.org/abs/2003.03917}, 
}

@misc{arun2025verdeverificationrefereeddelegation,
      title={Verde: Verification via Refereed Delegation for Machine Learning Programs}, 
      author={Arasu Arun and Adam St. Arnaud and Alexey Titov and Brian Wilcox and Viktor Kolobaric and Marc Brinkmann and Oguzhan Ersoy and Ben Fielding and Joseph Bonneau},
      year={2025},
      eprint={2502.19405},
      archivePrefix={arXiv},
      primaryClass={cs.LG},
      url={https://arxiv.org/abs/2502.19405}, 
}

@misc{durvasula2025resonance,
  title={Resonance: A Market Mechanism for Heterogeneous Computation},
  author={Naveen Durvasula and Maryam Bahrani},
  howpublished={\url{https://ritual.net/blog/resonance-pt1}},
  year={2025}
}

@misc{ritual2025chain,
  title={Introducing Ritual Chain},
  author={{Ritual Foundation}},
  howpublished={\url{https://ritualfoundation.org/blog/unveiling-ritual}},
  year={2025}
}

@INPROCEEDINGS{8751419,
  author={Chenli, Changhao and Li, Boyang and Shi, Yiyu and Jung, Taeho},
  booktitle={2019 IEEE International Conference on Blockchain and Cryptocurrency (ICBC)}, 
  title={Energy-recycling Blockchain with Proof-of-Deep-Learning}, 
  year={2019},
  volume={},
  number={},
  pages={19-23},
  doi={10.1109/BLOC.2019.8751419}}

@article{zheng2018blockchain,
  title={Blockchain challenges and opportunities: A survey},
  author={Zheng, Zibin and Xie, Shaoan and Dai, Hong-Ning and Chen, Xiangping and Wang, Huaimin},
  journal={International journal of web and grid services},
  volume={14},
  number={4},
  pages={352--375},
  year={2018},
  publisher={Inderscience Publishers (IEL)}
}

@article{liu2021proof,
  title={Proof of Learning (PoLe): Empowering neural network training with consensus building on blockchains},
  author={Liu, Yuan and Lan, Yixiao and Li, Boyang and Miao, Chunyan and Tian, Zhihong},
  journal={Computer Networks},
  volume={201},
  pages={108594},
  year={2021},
  publisher={Elsevier}
}

@article{baldominos2019coin,
  title={Coin. AI: A proof-of-useful-work scheme for blockchain-based distributed deep learning},
  author={Baldominos, Alejandro and Saez, Yago},
  journal={Entropy},
  volume={21},
  number={8},
  pages={723},
  year={2019},
  publisher={MDPI}
}

@article{li2019dlbc,
  title={Dlbc: A deep learning-based consensus in blockchains for deep learning services},
  author={Li, Boyang and Chenli, Changhao and Xu, Xiaowei and Shi, Yiyu and Jung, Taeho},
  journal={arXiv preprint arXiv:1904.07349},
  year={2019}
}

@article{zhi2025blockchain,
  title={Blockchain consensus scheme based on the proof of distributed deep learning work},
  author={Zhi, Hui and Wu, HongCheng and Huang, Yu and Tian, ChangLin and Wang, SuZhen},
  journal={IET Software},
  volume={2025},
  number={1},
  pages={3378383},
  year={2025},
  publisher={Wiley Online Library}
}

@article{moosavi2024blockchain,
  title={Blockchain technology, structure, and applications: a survey},
  author={Moosavi, Nazanin and Taherdoost, Hamed and Mohamed, Nachaat and Madanchian, Mitra and Farhaoui, Yousef and Khan, Inam Ullah},
  journal={Procedia Computer Science},
  volume={237},
  pages={645--658},
  year={2024},
  publisher={Elsevier}
}

@article{haouari2022novel,
  title={A novel proof of useful work for a blockchain storing transportation transactions},
  author={Haouari, Mohamed and Mhiri, Mariem and El-Masri, Mazen and Al-Yafi, Karim},
  journal={Information Processing \& Management},
  volume={59},
  number={1},
  pages={102749},
  year={2022},
  publisher={Elsevier}
}

\end{document}